\documentclass[aps,prd,twocolumn,amsmath,showpacs,superscriptaddress,nofootinbib,nopreprintnumbers,tightenlines]{revtex4-1}

\usepackage{verbatim}
\usepackage[T1]{fontenc}
\usepackage[utf8]{inputenc}
\usepackage[american]{babel}
\usepackage{epsfig}
\usepackage{graphicx}

\usepackage{hyperref}

\usepackage{booktabs}
\usepackage{multirow}
\usepackage{dcolumn}
\usepackage{amsmath}
\usepackage{mathtools}
\usepackage{amsfonts}
\usepackage{amssymb}
\usepackage{epstopdf}
\usepackage{bm}
\usepackage{siunitx}
\usepackage{braket}
\usepackage{enumitem}
\usepackage{soul}
\usepackage[table]{xcolor}
\usepackage{color}
\usepackage{transparent}
\usepackage{pifont}

\definecolor{navyblue}{rgb}{0.0, 0.0, 0.5}
\definecolor{royalblue}{rgb}{0.25, 0.41, 0.88}
\definecolor{cadmiumgreen}{rgb}{0.0, 0.42, 0.24}
\definecolor{blue-violet}{rgb}{0.54, 0.17, 0.89}
\definecolor{darkviolet}{rgb}{0.58, 0.0, 0.83}
\definecolor{teal(colorwheel)}{rgb}{1.0, 0.5, 0.0}

\usepackage{hyperref}
\hypersetup{
    colorlinks=true, 
    linkcolor=royalblue, 
    citecolor=magenta}

\newcommand\ee{\end{equation}}
\newcommand\be{\begin{equation}}
\newcommand\eea{\end{eqnarray}}
\newcommand\bea{\begin{eqnarray}}

\usepackage{booktabs}
\usepackage{multirow}
\usepackage{dcolumn}
\usepackage{colortbl}

\definecolor{magenta(process)}{rgb}{1.0, 0.0, 0.56}

\definecolor{darkspringgreen}{rgb}{0.09, 0.45, 0.27}

\definecolor{royalblue(web)}{rgb}{0.25, 0.41, 0.88}
\begin{document}

\title{Neutrino Mass Bounds in the era of Tension Cosmology}

\author{Eleonora Di Valentino}
\email{e.divalentino@sheffield.ac.uk}
\affiliation{School of Mathematics and Statistics, University of Sheffield, Hounsfield Road, Sheffield S3 7RH, United Kingdom}

\author{Alessandro Melchiorri}
\email{alessandro.melchiorri@roma1.infn.it}
\affiliation{Physics Department and INFN, Universit\`a di Roma ``La Sapienza'', Ple Aldo Moro 2, 00185, Rome, Italy}

\date{\today}


\begin{abstract}
The measurements of Cosmic Microwave Background anisotropies made by the Planck satellite provide extremely tight upper bounds on the total neutrino mass scale ($\Sigma m_{\nu}<0.26 eV$ at $95\%$ C.L.). However, as recently discussed in the literature, the Planck data show anomalies that could affect this result. Here we provide new constraints on neutrino masses using the recent and complementary CMB measurements from the Atacama Cosmology Telescope DR4 and the South Polar Telescope SPT-3G experiments.
We found that both the ACT-DR4 and SPT-3G data, when combined with WMAP, mildly suggest a neutrino mass with $\Sigma m_{\nu}=0.68 \pm 0.31$~eV and $\Sigma m_{\nu}=0.46_{-0.36}^{+0.14}$~eV at $68 \%$ C.L, respectively. Moreover, when CMB lensing from the Planck experiment is included, the ACT-DR4 data now indicates a neutrino mass above the two standard deviations, with $\Sigma m_{\nu}=0.60_{-0.50}^{+0.44}$~eV at $95 \%$, while WMAP+SPT-3G provides a weak upper limit of $\Sigma m_{\nu}<0.37$~eV at $68 \%$ C.L.. Interestingly, these results are consistent with the Planck CMB+Lensing constraint of $\Sigma m_{\nu} = 0.41_{-0.25}^{+0.17}$~eV at $68 \%$ C.L. when
variation in the $A_{\rm lens}$ parameter are considered. We also show that these indications are still present after the inclusion of BAO or SN-Ia data in extended cosmologies that are usually considered to solve the so-called Hubble tension. In this respect, we note that in these models CMB+BAO constraints prefer a higher neutrino mass for higher values of the Hubble constant. 
A combination of ACT-DR4, WMAP, BAO and constraints on the Hubble constant from the SH0ES collaboration gives $\Sigma m_{\nu}=0.39^{+0.13}_{-0.25}$~eV at $68 \%$ C.L. in extended cosmologies.
We conclude that a total neutrino mass above the $0.26$~eV limit still provides an excellent fit 
to several cosmological data and that future data must be considered before safely ruling it out. 
\end{abstract}

\maketitle

\section{Introduction}

It has been extensively shown that Cosmic Microwave Background anisotropies alone can provide a clean and robust constraint on the neutrino global mass scale $\Sigma m_{\nu}$ (see e.g.~\cite{Lesgourgues:2006nd,Gonzalez-Garcia:2007dlo,TopicalConvenersKNAbazajianJECarlstromATLee:2013bxd,Kaplinghat:2003bh,Pascoli:2005zb,Lattanzi:2017ubx, DeSalas:2018rby}). The Planck experiment, in combination with atmospheric and solar neutrino oscillation experiments, provides at the moment no indication for a neutrino mass with a reported limit on the sum of the three active neutrinos of $\Sigma m_{\nu} <0.26$~eV at $95 \%$ C.L.~\cite{Planck:2018vyg} from CMB angular spectra data. This constraint is further improved to $\Sigma m_{\nu}<0.12$~eV at $95 \%$ C.L. when Baryon Acoustic Oscillation (BAO) data are included~\cite{Planck:2018vyg}. A recent combination of Planck data with Supernovae Ia luminosity distances, BAO and determinations of the growth rate parameter, set the most constraining bound to date, $\Sigma m_{\nu} <0.09$~eV at $95\%$ C.L.~\cite{DiValentino:2021hoh}.

These strong limits have obviously important consequences for current and planned laboratory experiments devoted to neutrino mass detection (see, e.g.~\cite{Capozzi:2021fjo}). We remind the reader that an effective neutrino mass $m_{\beta}$ can be measured through beta-decay experiments, while an effective mass $m_{\beta \beta}$ can be obtained from neutrinoless double beta-decay experiments ($0\nu\beta\beta$) if neutrino are Majorana fermions. The Planck limit, in combination with neutrino oscillation data, suggest values of $m_{\beta}$ and $m_{\beta \beta}$ below the $100$~meV (see, e.g.~\cite{Capozzi:2021fjo}). These values are clearly challenging for current $0\nu\beta\beta$ experiments and clearly out of the reach of sensitivity of KATRIN, the on-going, state-of-the art, beta-decay experiment~\cite{KATRIN:2019yun}.

It is, however, also well known that the cosmological limits could be plagued by the so-called lensing anomaly present in the Planck angular spectra data~\cite{Planck:2018vyg,DiValentino:2019dzu}. Planck power spectra are indeed suggesting a larger gravitational lensing amplitude (described by the
$A_{\rm lens}$ parameter as defined in~\cite{Calabrese:2008rt}) from dark matter fluctuations than expected at about $99 \%$ C.L.. Since gravitational lensing anti-correlates with a neutrino component that prevent clustering, an anomalous higher value for $A_{\rm lens}$, as seen in the Planck data, can bias the neutrino mass constraints towards lower values~\cite{Capozzi:2021fjo}. 
While at the moment it is not clear if the $A_{\rm lens}$ anomaly is due to a systematic error or new physics, its presence suggests that Planck limits on neutrino masses should be taken with some grain of salt.

A possible solution to the problem, as we point out in this {\it letter}, comes from the new and exquisite CMB measurements provided by the Atacama Cosmology Telescope (ACT Data Release 4, ACT-DR4)~\cite{ACT:2020gnv} and the South Pole Telescope (SPT-3G)~\cite{SPT-3G:2021eoc}.  ACT-DR4 and SPT-3G, when combined with previous data from the WMAP satellite~\cite{WMAP:2012nax}, can provide limits on cosmological parameters with a constraining power comparable to Planck. These experiments show no $A_{\rm lens}$ anomaly and are therefore ideal for constraining neutrino masses and double-checking the Planck constraints. It is therefore extremely timely to revisit the cosmological constraints on neutrino masses using these new CMB data.

There is, however, one caveat. While both ACT-DR4 and SPT-3G are consistent with a standard lensing amplitude, they are less consistent with other standard expectations. The ACT-DR4 release, in particular, seem to suggest a neutrino effective number $N_{\rm eff}$ lower than $3.04$~\cite{ACT:2020gnv}, a running of the spectral index $dn/dlnk<0$ at about one-two standard deviations~\cite{ACT:2020gnv,Forconi:2021que}, and an Early Dark Energy at more than 99\% C.L.~\cite{Hill:2021yec,Poulin:2021bjr}. The SPT-3G data, on the other hand, appears more in agreement with a higher value for $N_{\rm eff}$~\cite{SPT-3G:2021eoc}. These small anomalies could also correlate with the constraints on neutrino masses and it is therefore important to perform an analysis in an extended parameter space (that includes also variations in $N_{\rm eff}$ and $dn/dlnk$) to check the model dependence of the neutrino constraints. Considering an extended parameter space with respect to the $6$ parameters LCDM model is also useful in view of the reported tensions present on the values of the Hubble constant~\cite{Verde:2019ivm,DiValentino:2020zio,DiValentino:2021izs,Perivolaropoulos:2021jda,Freedman:2021ahq,Shah:2021onj} and the $S_8$ parameter~\cite{DiValentino:2020vvd,Perivolaropoulos:2021jda} between CMB and local observables.

In what follows we will therefore use two approaches. Firstly we present the constraints on neutrino masses under the assumption of a LCDM scenario, while as a second step we also consider a more general framework where many additional parameters are considered, following the approach used in~\cite{DiValentino:2015ola,DiValentino:2016hlg,DiValentino:2017zyq,DiValentino:2019dzu,DiValentino:2020hov}.

This {\it letter} is structured in the following way: in Section~\ref{methods} we describe the datasets and the method used to extract the cosmological constraints, in Section~\ref{results} we present the results obtained for the different cases, and finally in Section~\ref{conclusions} we derive our conclusions.

\section{Method and datasets}
\label{methods}

In this section we describe the methodology used in our analysis and the cosmological datasets used to derive our results.
Our data analysis method follows the same procedure already used in several previous
papers for the CMB data. In practice the constraints on cosmological parameters are obtained adopting a Monte Carlo Markov Chain (MCMC) algorithm, using the public available CosmoMC package~\cite{Lewis:1999bs,Lewis:2002ah}, and the convergence of the chains is tested using the Gelman-Rubin criterion~\cite{Gelman:1992zz}. 

Regarding the datasets considered we have instead:

\begin{itemize}
    \item {\bf Planck}: Planck 2018 temperature and polarization anisotropies angular power spectra {\it plikTTTEEE+lowl+lowE} from the legacy release~\citep{Planck:2018vyg,Planck:2018nkj,Planck:2019nip};
    
    \item {\bf ACT-DR4}: Atacama Cosmology Telescope DR4 likelihood~\cite{ACT:2020gnv}, considering the multipoles $\ell>600$ in TT and $\ell>350$ in TE and EE;

    \item {\bf SPT-3G}: South Pole Telescope polarization measurements SPT-3G~\cite{SPT-3G:2021eoc}, considering the multipoles $300<\ell<3000$ in TE and EE;
    
    \item {\bf WMAP}: WMAP 9 years observations data~\cite{WMAP:2012nax}, considering the multipoles $20<\ell<1200$ in TT and $24<\ell<800$ in TE;
    
    \item {\bf Lensing}: Planck 2018 lensing reconstruction power spectrum obtained from the CMB trispectrum analysis~\cite{Planck:2018lbu};
    
    \item {\bf tauprior}: a gaussian prior on the optical depth $\tau = 0.065 \pm 0.015$, as used in~\cite{ACT:2020gnv};

    \item {\bf BAO}: Baryon Acoustic Oscillations measurements from 6dFGS~\citep{Beutler:2011hx}, SDSS MGS~\cite{Ross:2014qpa} and BOSS DR12~\cite{BOSS:2016wmc} surveys;
    
    \item {\bf Pantheon}: Pantheon sample~\cite{Scolnic:2017caz} of the Type Ia Supernovae, consisting of 1048 data points, distributed in the redshift interval \mbox{$0.01 \leq z \leq 2.3$}.
    
    \item {\bf R20}: A gaussian prior on the Hubble constant as measured by the SH0ES collaboration~\cite{Riess:2020fzl}, i.e. $H_0=73.2\pm1.3$~km/s/Mpc.
     
    \item {\bf F21}: A gaussian prior on the Hubble constant as measured by Freedman in her review~\cite{Freedman:2021ahq}, i.e. $H_0=69.8\pm1.7$~km/s/Mpc.
    
\end{itemize}

Our baseline parameter space consists of the 6 parameters of the $\Lambda$CDM model (baryon $\Omega_{\rm b}h^2$ and cold dark matter $\Omega_{\rm c}h^2$ energy densities, angular size of the horizon at the last scattering surface $\theta_{\rm{MC}}$, optical depth $\tau$, amplitude and spectral index of primordial scalar perturbation $A_{\rm s}$ and $n_s$) plus a total neutrino mass $\Sigma m_{\nu}$ free to vary (i.e. 7 parameters in total), but we consider also a second 10 parameters scenario where we include at the same time the effective neutrino number $N_{\rm eff}$, the dark energy equation of state $w$, and the running of the scalar spectral index $dn/dlnk$. Only when the Planck data are included in the analysis, we consider also variations in the $A_{\rm lens}$ parameter, in order to marginalize over the anomaly (or possible systematic error) present in the Planck data, i.e. we will have 8 parameters varying in the baseline case and 11 parameters in the extended case.
For all the cosmological parameters varied in our analysis we choose flat prior distributions as listed in Table~\ref{Priors}. 

\begin{table}
	\begin{center}
		\renewcommand{\arraystretch}{1.5}
		\begin{tabular}{c@{\hspace{0. cm}}@{\hspace{1.5 cm}} c}
			\hline
			\textbf{Parameter}    & \textbf{Prior} \\
			\hline\hline
			$\Omega_{\rm b} h^2$         & $[0.005\,,\,0.1]$ \\
			$\Omega_{\rm c} h^2$     	 & $[0.001\,,\,0.99]$\\
			$100\,\theta_{\rm {MC}}$     & $[0.5\,,\,10]$ \\
			$\tau$                       & $[0.01\,,\,0.8]$\\
			$\log(10^{10}A_{\rm S})$     & $[1.61\,,\,3.91]$ \\
			$n_{\rm s}$                  & $[0.8\,,\, 1.2]$ \\
			$\Sigma m_{\nu}$ [eV]             & $[0.06\,,\, 5]$ \\ 
			$w$              & $[-3\,,\, 1]$ \\
			$N_{\rm eff}$                          & $[0.05\,,\, 10]$ \\	
		    $dn/dlnk$             & $[-1\,,\,1]$\\
		    $A_{\rm lens}$             & $[0\,,\,10]$\\
			\hline\hline
		\end{tabular}
		\caption{List of the parameter priors.}
		\label{Priors}
	\end{center}
\end{table}

\begin{table}
\begin{center}
		\renewcommand{\arraystretch}{1.5}
\begin{tabular}{c|c}
\hline \hline
~~~~~~~~~Dataset~~~~~~~~~                    & ~~~~~~~~~$\Sigma m_{\nu}$ [eV]~~~~~~~~~\\
\hline 
Planck (+$A_{\rm lens}$)            & $<0.51$\\
Planck+BAO  (+$A_{\rm lens}$)             & $<0.19$\\
Planck+Pantheon  (+$A_{\rm lens}$)        & $<0.25$\\
Planck+Lensing   (+$A_{\rm lens}$)       & $0.41^{+0.17}_{-0.25}$\\
\hline 
ACT-DR4+WMAP             & $0.68\pm0.31$\\
ACT-DR4+WMAP+BAO             & $<0.19$\\
ACT-DR4+WMAP+Pantheon        & $<0.25$\\
ACT-DR4+WMAP+Lensing                 & $0.60\pm0.25$\\
\hline
SPT-3G+WMAP            & $0.46^{+0.14}_{-0.36}$\\
SPT-3G+WMAP+BAO             & $0.22^{+0.056}_{-0.14}$\\
SPT-3G+WMAP+Pantheon        & $0.25^{+0.052}_{-0.19}$\\
SPT-3G+WMAP+Lensing                        & $<0.37$\\
\hline
\hline
\hline
\end{tabular}
\end{center}
\caption{Constraints on the sum of neutrino masses $\Sigma m_{\nu}$ at $68 \%$ C.L. from
a combination of different datasets in case of the LCDM+$\Sigma m_{\nu}$ scenario. A prior on the optical depth (tauprior) is included in all the analysis involving the ACT and SPT datasets.
\label{tab:nulimits}
}
\end{table}

\begin{figure*}
\includegraphics[width=0.4\textwidth]{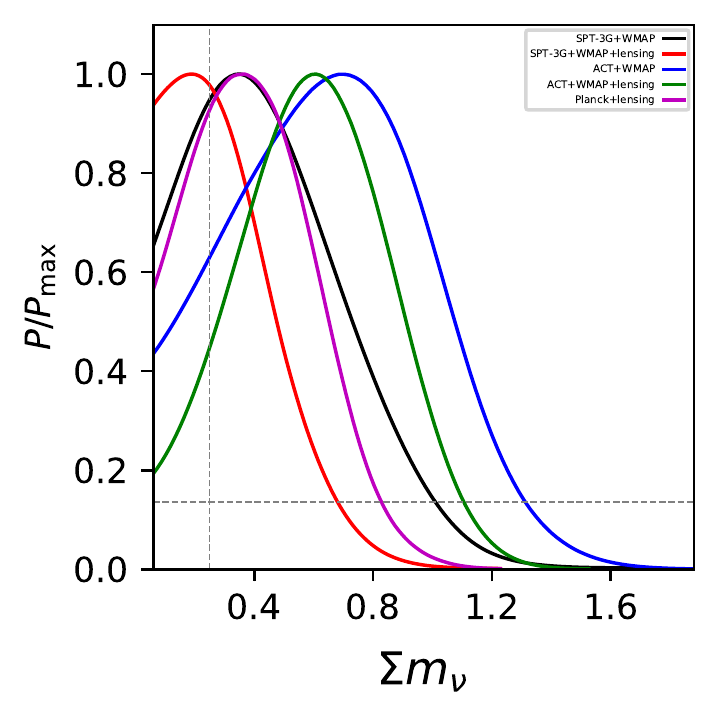}
\includegraphics[width=0.4\textwidth]{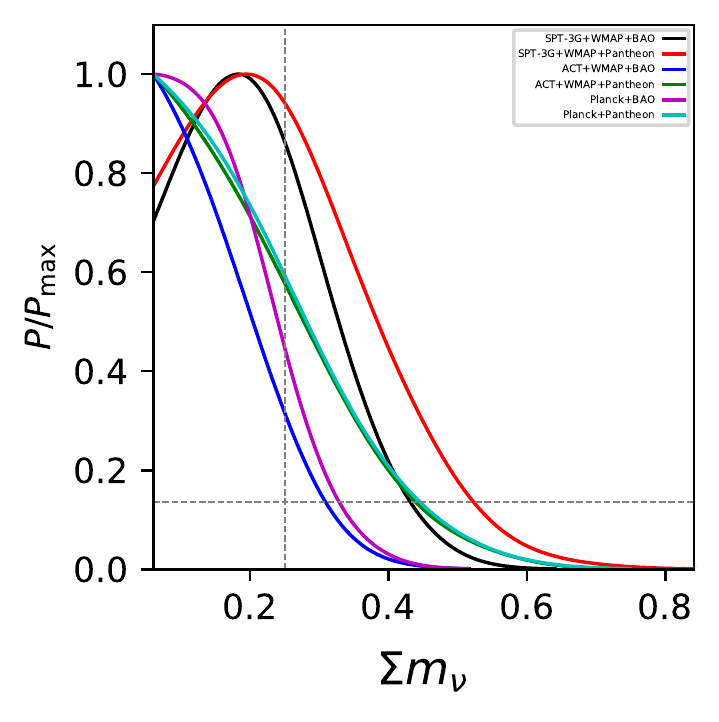}
\caption{One dimensional posterior distributions for $\Sigma m_{\nu}$ for several combination
of datasets under the assumption of $\Lambda$CDM model (Planck analysis includes variation in the $A_{\rm lens}$ parameter). A prior on the optical depth (tauprior) is present every time the ACT-DR4 and SPT-3G data are considered, but neglected in the legend for brevity. The vertical dashed line identifies the value $\Sigma m_{\nu}=0.26$~eV while
the horizontal line is for $P=0.135$, i.e. where a Gaussian distribution is at two standard deviation from the maximum. None of the posteriors exclude a value of $\Sigma m_{\nu}=0.26$~eV.}
\label{figlcdm}
\end{figure*}

\begin{table}
\begin{center}
		\renewcommand{\arraystretch}{1.3}
\begin{tabular}{c|c}
\hline \hline
~~~~~~~~~Dataset~~~~~~~~~                    & ~~~~~~~~~$\Sigma m_{\nu}$ [eV]~~~~~~~~~\\
\hline 
Planck (+$A_{\rm lens}$)            & $<0.50$\\
Planck+BAO  (+$A_{\rm lens}$)             & $<0.22$\\
Planck+Pantheon  (+$A_{\rm lens}$)        & $<0.47$\\
Planck+Lensing   (+$A_{\rm lens}$)        & $0.38^{+0.12}_{-0.28}$\\
\hline 
ACT-DR4+WMAP             & $0.81\pm0.28$\\
ACT-DR4+WMAP+BAO             & $<0.27$\\
ACT-DR4+WMAP+Pantheon        & $0.71\pm0.28$\\
ACT-DR4+WMAP+Lensing       & $0.56\pm0.21$\\
ACT-DR4+WMAP+R20       & $0.83\pm0.230$\\
ACT-DR4+WMAP+F21       & $0.85^{+0.27}_{-0.33}$\\
ACT-DR4+WMAP+BAO+R20             & $0.39^{+0.13}_{-0.25}$\\
ACT-DR4+WMAP+BAO+F21             & $<0.34$\\
\hline
SPT-3G+WMAP             & $<0.56$\\
SPT-3G+WMAP+BAO             & $<0.28$\\
SPT-3G+WMAP+Pantheon        & $0.46^{+0.11}_{-0.39}$\\
SPT-3G+WMAP+Lensing                        & $<0.39$\\
SPT-3G+WMAP+R20                        & $0.49^{+0.12}_{-0.42}$\\
SPT-3G+WMAP+F21                       & $<0.60$\\
SPT-3G+WMAP+BAO+R20             & $0.37^{+0.13}_{-0.25}$\\
SPT-3G+WMAP+BAO+F21             & $<0.32$\\
\hline

\hline
\hline
\end{tabular}
\end{center}
\caption{Constraints on the sum of neutrino masses $\Sigma m_{\nu}$ at $68 \%$ C.L. from
a combination of different datasets in case of the LCDM+$\Sigma m_{\nu}$+$w$+$N_{\rm eff}$+$dn/dlnk$ scenario. A prior on the optical depth (tauprior) is included in all the analysis involving the ACT and SPT datasets.
\label{tab:nucosmolimits}
}
\end{table}

\begin{figure*}
\includegraphics[width=0.4\textwidth]{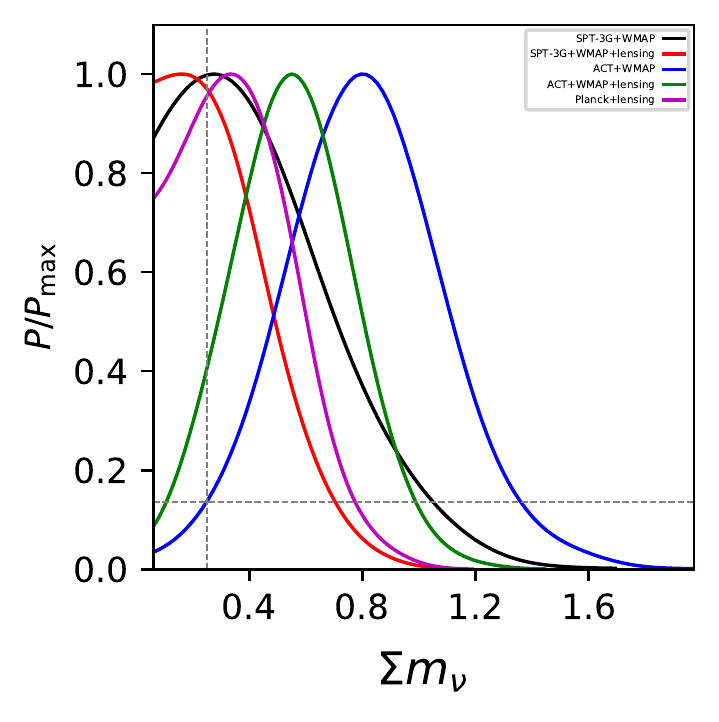}
\includegraphics[width=0.4\textwidth]{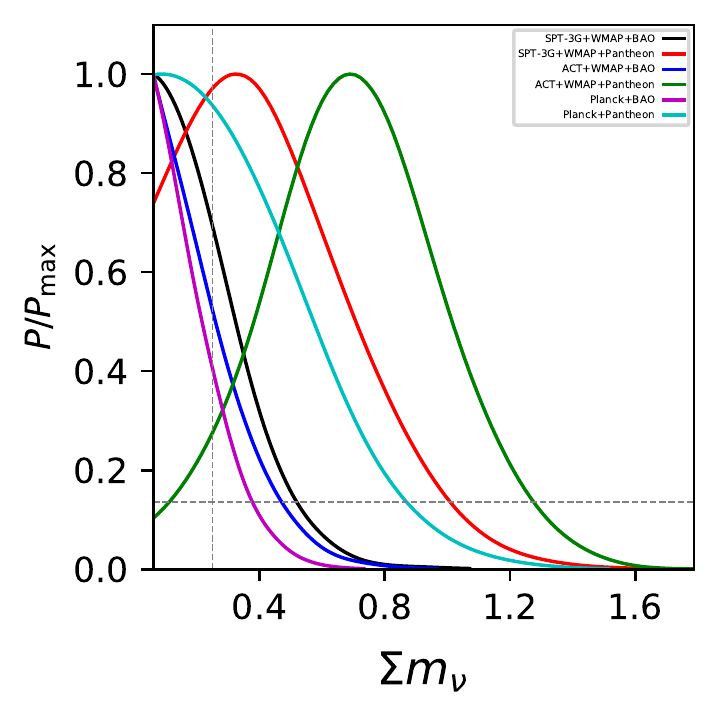}
\caption{One dimensional posterior distributions for $\Sigma m_{\nu}$ for several combination
of datasets under the assumption of an extended $\Lambda$CDM model (Planck analysis includes variation in the $A_{\rm lens}$ parameter). A prior on the optical depth (tauprior) is present every time the ACT-DR4 and SPT-3G data are considered, but neglected in the legend for brevity. The vertical dashed line identifies the value $\Sigma m_{\nu}=0.26$~eV while the horizontal line is for $P=0.135$, i.e. where a Gaussian distribution is at two standard deviation from the maximum. None of the posteriors exclude a value of $\Sigma m_{\nu}=0.26$~eV.}
\label{figextlcdm}
\end{figure*}

\begin{figure*}
\includegraphics[width=0.32\textwidth]{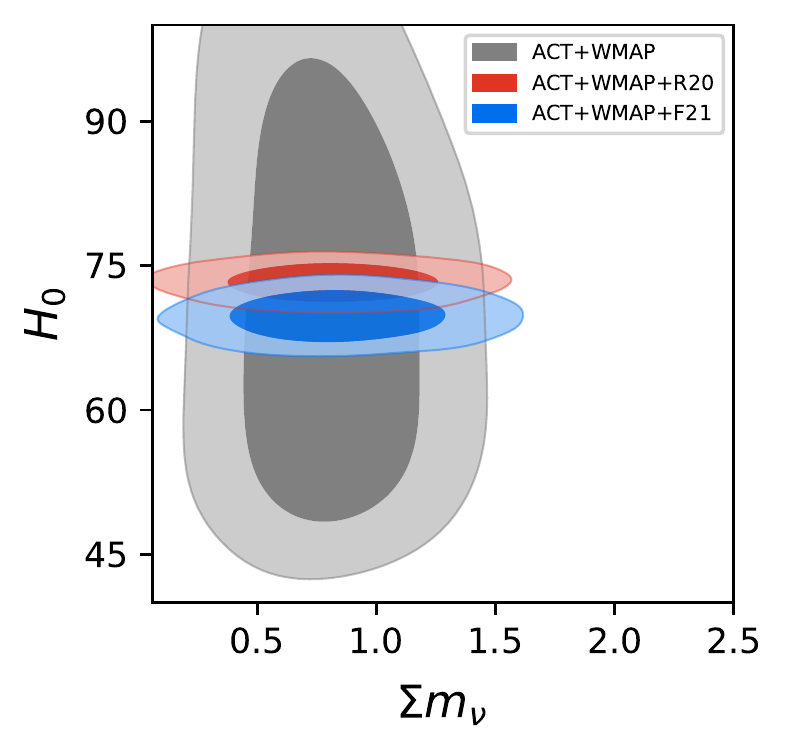}
\includegraphics[width=0.32\textwidth]{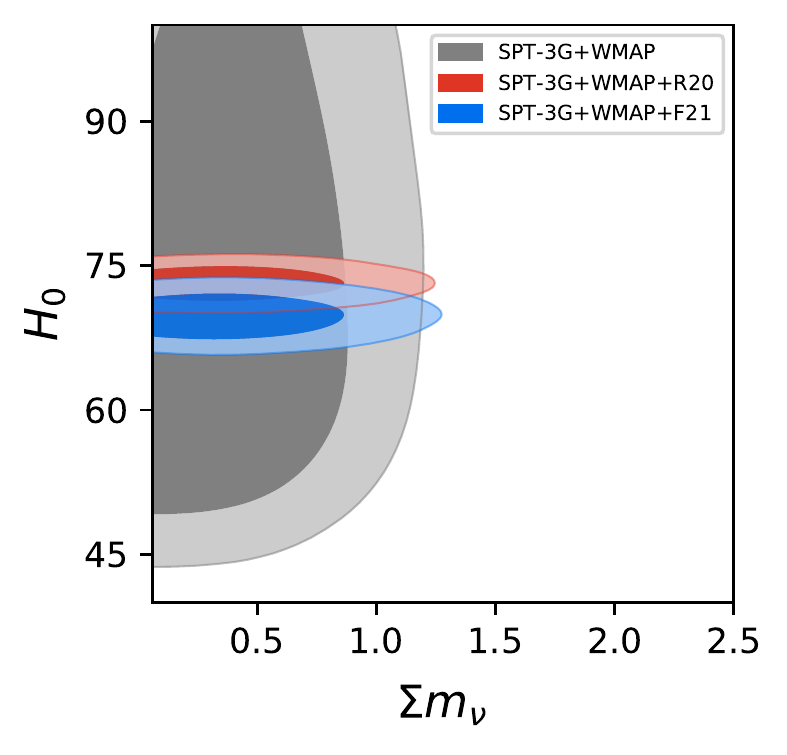}
\includegraphics[width=0.32\textwidth]{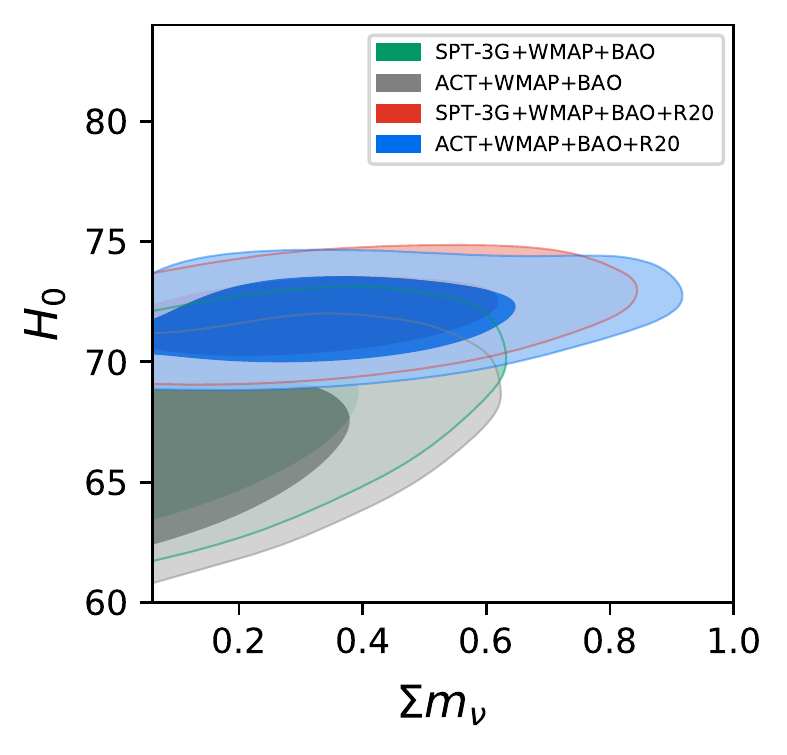}
\caption{2D contour plots for $\Sigma m_{\nu}$ vs $H_0$ for the ACT-DR4 (left panel) and SPT-3G (middle panel) combinations with WMAP and the gaussian priors on the Hubble constant R20 and F21, (a prior on the optical depth (tauprior) is present everywhere, but neglected in the legend for brevity), and with the addition of BAO (right panel), under the assumption of an extended $\Lambda$CDM model.}
\label{fig2D}
\end{figure*}

\section{Results}
\label{results}

In this section we describe the results we obtained with different dataset combinations in the $\Lambda$CDM+$\Sigma m_{\nu}$ case and in the extended $\Lambda$CDM scenario with multi parameters free to vary.

\subsection{$\Lambda$CDM+$\Sigma m_{\nu}$ case.}

We report the constraints on the sum of neutrino masses $\Sigma m_{\nu}$ obtained under the assumption
of $\Lambda$CDM+$\Sigma m_{\nu}$ in Table~\ref{tab:nulimits} and in Figure~\ref{figlcdm}. The constraints obtained with the use of the Planck dataset assume also a variation in the $A_{\rm lens}$ parameter. 
As one can see we found no combination of datasets that can exclude a value of $\Sigma m_{\nu} >0.26$~eV at more than $95 \%$ C.L..

In particular we have that both ACT-DR4 and SPT-3G data, in their combination with WMAP and the tauprior, in order to be "Planck-independent", mildly suggest a neutrino mass with $\Sigma m_{\nu}=0.68 \pm 0.31$~eV and $\Sigma m_{\nu}=0.46_{-0.36}^{+0.14}$~eV at $68 \%$ C.L., respectively. If we then include the Lensing dataset from the Planck experiment the ACT-DR4+WMAP+tauprior+Lensing data prefers a neutrino mass above two standard deviations, with $\Sigma m_{\nu}=0.60\pm 0.25$~eV at $68 \%$ C.L., while SPT-3G+WMAP+tauprior+Lensing provides a weak upper limit of $\Sigma m_{\nu}<0.37$~eV at $68 \%$ C.L.. Interestingly, these results are well consistent with the Planck+Lensing constraint of $\Sigma m_{\nu} = 0.41_{-0.25}^{+0.17}$~eV at $68 \%$ C.L. when variation in the $A_{\rm lens}$ parameter are considered. 

We can also notice that these indications are not in strong disagreement with the data after the inclusion of BAO or Pantheon measurements.
As expected, the more stringent constraints are obtained when CMB and BAO data are combined together. However, while Planck+BAO and ACT-DR4+WMAP+tauprior+BAO exclude $\Sigma m_{\nu} >0.26$~eV just by $\sim 1.3\sigma$, the combination of SPT-3G+WMAP+tauprior+BAO shows a mild preference for $\Sigma m_{\nu} \sim 0.22$~eV
at about just one standard deviation. Interestingly, all the other data combinations, including those with Pantheon, show a value of $\Sigma m_{\nu} =0.26$~eV well inside one standard deviation, or even prefer a neutrino mass above this limit. 
While there is no clear indication for a neutrino mass, we can conclude also that there is no
data combination that could strongly disfavour a neutrino mass such that $\Sigma m_{\nu} =0.26$~eV.

\subsection{Extended $\Lambda$CDM.}

Given the current tensions between the cosmological datasets we also perform an analysis in an extended
$\Lambda$CDM scenario. Beside the usual $6$ parameters of the $\Lambda$CDM model we also consider
variations in the effective neutrino number $N_{\rm eff}$, the dark energy equation of state $w$, and the running of
the scalar spectral index $dn/dlnk$. The results are reported in Table~\ref{tab:nucosmolimits} and the posterior distributions shown in Figure~\ref{figextlcdm}.
The first thing we can notice is that the constraints on the neutrino mass are only slightly affected in this extended scenario with respect
to the $\Lambda$CDM+$\Sigma m_{\nu}$ case.

In particular now ACT-DR4 + WMAP + tauprior suggests a neutrino mass with $\Sigma m_{\nu}=0.81 \pm 0.28$~eV at 68\% C.L., shifted towards higher values but with comparable error bars with respect to the $\Lambda$CDM+$\Sigma m_{\nu}$ case. On the contrary, SPT-3G + WMAP + tauprior has now only an upper limit $\Sigma m_{\nu}<0.56$~eV at 68\% C.L..
The inclusion of the Lensing dataset gives instead for ACT-DR4 + WMAP + tauprior + Lensing a total neutrino mass bound of $\Sigma m_{\nu}=0.56\pm 0.21$~eV at $68 \%$ C.L., and for SPT-3G + WMAP + tauprior + Lensing a weak upper limit of $\Sigma m_{\nu}<0.39$~eV at $68 \%$ C.L., i.e. very similar to the bounds obtained in Table~\ref{tab:nulimits}.
Also in this extended scenario these results are in agreement with Planck+Lensing that finds $\Sigma m_{\nu} = 0.38_{-0.28}^{+0.12}$~eV at $68 \%$ C.L. where $A_{\rm lens}$ is free to vary. 

In this scenario, also when CMB and BAO data are combined together the value of $\Sigma m_{\nu} =0.26$~eV is well inside one standard deviation.

Another interesting point is that ACT-DR4 + WMAP + tauprior and SPT-3G + WMAP + tauprior both give the Hubble constant almost unconstrained in this extended scenario. For this reason it is possible here to safely use a Gaussian prior on this parameter and evaluate the effect on the neutrino masses. We can notice that the inclusion of the R20 or F21 priors does not affect the total neutrino mass constraints, because $H_0$ and $\Sigma m_{\nu}$ do not show any correlation (see Figure~\ref{fig2D}). Moreover, in this case it is alleviated also the long standing $S_8$ tension~\cite{DiValentino:2020vvd} with the weak lensing measurements, and we find that ACT-DR4 + WMAP + tauprior + R20 gives $S_8=0.726\pm0.037$ at 68\% C.L., while SPT-3G + WMAP + tauprior + R20 gives $S_8=0.732\pm0.037$ at 68\% C.L..  

Interestingly, as we can see from the last plot on the right of Figure~\ref{fig2D}, when CMB and BAO constraints are considered in these extended cosmologies, they provide constraints on the $\Sigma m_{\nu}$ vs $H_0$ plane that clearly show a correlation between these two parameters (higher neutrino masses are in better agreement with higher values of the Hubble constant). This degeneracy is exactly the opposite of what is obtained under standard LCDM, where higher neutrino masses prefer lower values of the Hubble constant. Therefore, in extended cosmologies that could solve the Hubble tension~\cite{DiValentino:2016hlg,DiValentino:2019dzu,DiValentino:2020hov}, a neutrino mass is {\it preferred} by the cosmological data, as we can also see from the 
ACT-DR4+WMAP+tauprior+BAO+R20 and SPT-3G+WMAP+tauprior+BAO+R20 constraints, that are $\Sigma m_{\nu}=0.39^{+0.13}_{-0.25}$~eV and $\Sigma m_{\nu}=0.37^{+0.13}_{-0.25}$~eV at $68 \%$ C.L. from Table~\ref{tab:nucosmolimits}, respectively.

\section{Conclusions}
\label{conclusions}
In this {\it letter} we constrained the total neutrino mass making use of recent CMB experiments such as ACT-DR4 and SPT-3G, combined with WMAP and a prior on the optical depth. In this way we obtained a neutrino masses estimate that is independent of the Planck data that appears as affected by the lensing anomaly $A_{\rm lens}$, possibly originating from an undetected systematic error.
We found that both the ACT and SPT experiments are well compatible, and in some cases also mildly suggesting, a total neutrino mass larger than the standard value, usually assumed in the literature equal to $0.06$~eV. Interestingly, when the Lensing dataset from Planck is included in the data analysis, ACT-DR4 hints to a neutrino mass above the two standard deviations, with $\Sigma m_{\nu}=0.60_{-0.50}^{+0.44}$~eV at $95 \%$ C.L., that is perfectly consistent with the Planck CMB+Lensing constraint of $\Sigma m_{\nu} = 0.41_{-0.25}^{+0.17}$~eV at $68 \%$ C.L. when one marginalizes over the $A_{\rm lens}$ parameter.
This indication for massive neutrinos is still present after the inclusion of BAO or SN-Ia data in extended cosmologies, involving additional parameters like the effective neutrino number $N_{\rm eff}$, the dark energy equation of state $w$ and the running of
the scalar spectral index $dn/dlnk$. In this respect, we have noted that in these extended scenarios CMB+BAO constraints prefer a higher neutrino mass for higher values of the Hubble constant. A combination of ACT-DR4, WMAP, BAO and constraints on the Hubble constant from the SH0ES collaboration~\cite{Riess:2020fzl} gives $\Sigma m_{\nu}=0.39^{+0.13}_{-0.25}$~eV at $68 \%$ C.L. in extended cosmologies.

We conclude that a total neutrino mass above the $0.26$~eV limit still provides an excellent fit to several cosmological data, and that, in light of the current cosmological tensions, future cosmological data must be considered before safely ruling it out.


\begin{acknowledgments}

EDV is supported by a Royal Society Dorothy Hodgkin Research Fellowship. AM is supported by TASP, iniziativa specifica INFN.
\end{acknowledgments}

\bibliography{bibfiles.bib}


\vfill
\end{document}